\documentclass[11pt]{article}

\usepackage{authblk}
\usepackage{amsmath,amssymb,array}
\usepackage[utf8]{inputenc}
\usepackage[T1]{fontenc}
\usepackage{booktabs}
\usepackage{graphicx}
\usepackage{multirow}
\usepackage[round]{natbib}
\usepackage{bm}
\usepackage{physics}
\usepackage[ruled, linesnumbered]{algorithm2e}
\usepackage{subfigure}
\usepackage{hyperref}

\begin{document}
\sloppy

\title{\textbf{hdbayes}: An \texttt{R} Package for Bayesian Analysis of Generalized Linear Models Using Historical Data}

\author[1]{Ethan M. Alt\footnote{The first two authors contributed equally to this manuscript.}}
\author[1]{Xinxin Chen$^*$\thanks{xchen2@unc.edu}}
\author[2]{Luiz M. Carvalho}
\author[1]{Joseph G. Ibrahim}
\affil[1]{University of North Carolina at Chapel Hill}
\affil[2]{Getulio Vargas Foundation}

\date{June 21, 2025}

\maketitle

\begin{abstract}
There has been increased interest in the use of historical data to formulate informative priors in regression models. While many such priors for incorporating historical data have been proposed, adoption is limited due to access to software. Where software does exist, the implementations between different methods could be vastly different, making comparisons between methods difficult. In this paper, we introduce the \texttt{R} package \textbf{hdbayes}, an implementation of the power prior, normalized power prior, Bayesian hierarchical model, robust meta-analytic prior, commensurate prior, and latent exchangeability prior for generalized linear models. The bulk of the package is written in the Stan programming language, with user-friendly \texttt{R} wrapper functions to call samplers.
\end{abstract}

\newpage
\section{Introduction}

A common occurrence in practical applications of Bayesian modeling is the use of historical data to inform the prior for the analysis of the data set at hand, which we call the current data.
For example, in clinical trials, one may be in possession of a Phase III data set in addition to the Phase II trial already completed. In such cases, it is often desirable to elicit an informative prior for the regression coefficients based on the historical data.

Many methods for incorporating historical data have been proposed, including the power prior \citep{ibrahim2000power}, the normalized power prior \citep{duan2006evaluating,carvalho2021normalized}, the Bayesian hierarchical model, commensurate priors \citep{hobbs2012commensurate}, and robust meta-analytic predictive priors \citep{schmidli2014robust}. Unfortunately, software implementations of these methods may be difficult to come across.
Where they do exist, the implementations can differ so substantially that it becomes troublesome to utilize multiple packages to compare methods.
Furthermore, many current implementations may utilize Metropolis-type samplers, which can be difficult to tune.

The \texttt{R} package \textbf{hdbayes} \citep{alt2024hdbayes} aims to fill this gap. In particular, \textbf{hdbayes} offers user-friendly \texttt{R} functions to conduct Bayesian analysis for generalized linear models (GLMs) using historical data, with consistent syntax across methods. The backbone of the package is written in the Stan programming language \citep{carpenter2017stan}, implemented in the \textbf{cmdstanr} package \citep{cmdstanr_pkg}. Although \textbf{cmdstanr} is not available on the Comprehensive R Archive Network (CRAN), \textbf{hdbayes} uses the \textbf{instantiate} package \citep{landau2023instantiate}, which depends on \textbf{cmdstanr}, to create pre-compiled code. 

Stan utilizes a highly efficient Markov chain Monte Carlo (MCMC) method known as Hamiltonian Monte Carlo (HMC), which requires little-to-no tuning from the user's perspective. In particular, Stan implements a highly optimized variant of the No U-Turn Sampler (NUTS) algorithm \citep{hoffman2014no}. 

The remainder of this paper proceeds as follows. 
In Section \hyperref[sec:priorelicit]{2}, we review methods for prior elicitation using historical data that are implemented in the \textbf{hdbayes} package, indicating any existing publicly available implementations for each prior. 
We provide methodology and code examples for model selection via marginal likelihoods in Section \hyperref[sec:normconst]{3}. 
In Section \hyperref[sec:dataanalysis]{4}, we illustrate the utility of our package via analyses of real data sets in AIDS clinical trials, comparing posterior results across all implemented priors in \textbf{hdbayes}.
We close with some discussion in Section \hyperref[sec:discussion]{5}.

\section[Prior elicitation with historical data]{Prior elicitation with historical data} \label{sec:priorelicit}

In this section, we review the priors that are implemented in the \textbf{hdbayes} package.
Where applicable, we also discuss existing software implementations for each prior.
In particular, we focus on prior elicitation for generalized linear models (GLMs, \citet{mccullagh1989generalized}), whose likelihood function is given by
\begin{align}
    L(\bm{\beta}, \phi | \bm{y}, \bm{X}) \propto \prod_{i=1}^n \exp\left\{ \frac{1}{a_i(\phi)}\left[ y_i \theta_i - b(\theta_i) \right] + c(y_i, \phi_i) \right\},
    \label{eq:glm_likelihood}
\end{align}
where $\theta_i = \theta(\bm{x}_i'\bm{\beta})$, $\theta$ is referred to as the $\theta$-link function, $b$ and $c$ are determined by the probability distribution, $\bm{\beta}$ is a $p\times 1$-dimensional vector of regression coefficients corresponding to the $p \times 1$ vector of covariates $\bm{x}_i$ (where $\bm{X} = (\bm{x}_1, \ldots, \bm{x}_n)'$), both of which may include an intercept, $y_i$ is a response variable (and $\bm{y} = (y_1, \ldots, y_n)'$), and $\phi > 0$ is a dispersion parameter, which is known and equal to $1$ for binomial and Poisson models.
For ease of exposition, we assume $a_i(\phi) = \phi$. Note that GLMs are sometimes parameterized in terms of the mean of the $i^{th}$ individual, $\mu_i$, and the $\mu$-link function: $g(\mu_i) = \bm{x}_i'\bm{\beta}$.
In this case, $\theta(\cdot) = (\dot{b}^{-1} \circ g^{-1})(\cdot)$, where $\dot{f}$ denotes the first derivative of the function $f$.
When $a_i(\phi) = \phi$, the likelihood in \eqref{eq:glm_likelihood} may be written in matrix form as
\begin{align}
    L(\bm{\beta}, \phi | \bm{y}, \bm{X}) \propto
         \exp\left\{ \frac{1}{\phi} \left[ \bm{y}'\theta(\bm{X} \bm{\beta}) - \bm{1}_n' b(\theta(\bm{X}\bm{\beta})) \right]
         + \bm{1}_n' c(\bm{y}, \phi)
         \right\},
         \label{eq:glm_likelihood_matrix}
\end{align}
where $\bm{1}_n = (1, \ldots, 1)'$ is an $n\times 1$-dimensional vector of ones and the functions $\theta(\cdot)$, $b(\cdot)$, and $c(\cdot, \phi)$ are evaluated component-wise.

Let the current data set be denoted by $D = \{ (y_i, \bm{x}_i), i = 1, \ldots, n \}$, where $n$ denotes the sample size of the current data. Suppose we have $H$ historical data sets, and let the $h^{th}$ historical data set, be denoted by $D_{0h} = \{ (y_{0hi}, \bm{x}_{0hi}), i = 1, \ldots, n_{0h} \}$ for $h = 1, \ldots, H$, where $n_{0h}$ denotes the sample size of the $h^{th}$ historical data. Let $D_0 = \{D_{01}, \ldots, D_{0H}\}$ denote all the historical data.

In the sequel, we discuss the priors that are implemented in \textbf{hdbayes}. We compare with several other \texttt{R} packages that provide varying functionality for historical data priors.

\subsection{Basic syntax}
The \textbf{hdbayes} package was designed to enable users familiar with the \texttt{R} programming language (and, in particular, its \texttt{glm} function) to be able to use historical data borrowing priors. To that end, the \textbf{hdbayes} package has a basic level of syntax that is common to all implemented priors. In particular, each implemented prior takes, in pseudo-code the form \texttt{glm.prior(formula, family, data.list, prior.args, ...)}, where \texttt{formula} is a two-sided formula object, \texttt{family} is a family object containing a distribution-link function pair, \texttt{data.list} is a list of \texttt{data.frame}s, where the first element of the list is the current data set and all remaining elements are treated as historical data sets, \texttt{prior.args} is pseudo-code for a placeholder for arguments (e.g., hyperparmaeters) that are specific to the prior, and the ellipsis (\texttt{...}) contains arguments to pass onto the sampler in \textbf{cmdstanr} (e.g., the number of chains, the warm-up period, etc.). Note that the first two arguments of \texttt{glm.prior} are precisely the same as \texttt{glm}, while the third argument of \texttt{glm.prior} is a list of data frames while that in \texttt{glm} is a single data frame. As will be discussed in detail, each implemented prior in \textbf{hdbayes} has sensible default values for \texttt{prior.args}, making it easy for the end user to apply the methods. A summary of the functionality of \textbf{hdbayes} compared to other packages is presented in Table~\ref{tab:comparison}.

\renewcommand{\arraystretch}{1.3}
\begin{table}[htbp]
\centering
\resizebox{1\textwidth}{!}{%
\begin{tabular}{|l|c|c|c|c|c|}
\hline
  & \multicolumn{5}{c|}{\textbf{Package Name}} \\ \cline{2-6}
  & \textbf{hdbayes} & \textbf{NPP} & \textbf{BayesPPD} & \textbf{psborrow2} & \textbf{RBesT} \\ \hline
Power prior                           & X &   &   &   &       \\ \hline
Normalized power prior                & X & X & X &   &       \\ \hline
Normalized asymptotic power prior     & X &   &   &   &       \\ \hline
Robust meta-analytic predictive prior & X &   &   &   & X     \\ \hline
Bayesian hierarchical model           & X &   &   &   &       \\ \hline
Commensurate prior                    & X &   &   & X &       \\ \hline
LEAP                                  & X &   &   &   &       \\ \hline
All models in \texttt{stats::glm}     & X &   &   &   &       \\ \hline
$>1$ historical data set              & X & X & X &   &       \\ \hline
Marginal likelihood calculation       & X &   &   &   &       \\ \hline
\end{tabular}%
}
\caption{Features present in various \texttt{R} packages providing support for prior elicitation on the basis of historical data.
}
\label{tab:comparison}
\end{table}

\subsection{Power prior} 
\label{sec:pp}
The power prior (PP) of \citet{ibrahim2000power}, developed for single historical data set settings, involves discounting the likelihood of the historical data by some value $a_{01} \in [0, 1]$ (often referred to as the \emph{discounting parameter}) along with eliciting an initial prior $\pi_0$. We may express this mathematically as
\begin{align}
    \pi_{\text{PP}}(\bm{\beta}, \phi | D_{01}, a_{01}, \pi_0) 
    = \frac{L(\bm{\beta}, \phi | D_{01})^{a_{01}} \pi_0(\bm{\beta}, \phi)}{Z(a_{01})}
    \propto L(\bm{\beta}, \phi | D_{01})^{a_{01}} \pi_0(\bm{\beta}, \phi),
    \label{eq:pp_fixeda0_singledataset}
\end{align}
where $Z(a_{01}) = \int_{\mathbb{R}^p} \int_{0}^{\infty} L(\bm{\beta}, \phi | D_{01})^{a_{01}} \pi_0(\bm{\beta}, \phi) d\phi ~d\bm{\beta}$ is a normalizing constant, whose value is unimportant when $a_{01}$ is fixed.

For fixed $a_{01}$, the effective sample size of the PP (i.e., the number of observations that the prior is `worth') is given by $a_{01} n_{01}$, which is easily computable. The initial prior $\pi_0$ is typically taken to be non-informative since the goal is for the prior to depend heavily on the historical data. Note that when $a_{01} = 0$ the PP is the initial prior, and when $a_{01} = 1$ the PP is the posterior of the historical data. The PP thus provides a flexible way to incorporate historical data and quantify the informativeness of the prior. \citet{ibrahim2015power} provides an overview of how to select $a_{01}$.
In general, it is recommended to try several choices of $a_{01}$ to see how sensitive the posterior is to the value of $a_{01}$.

In \textbf{hdbayes}, we extend the traditional PP to accommodate multiple historical data sets by allowing users to elicit a vector of discounting parameters $\bm{a}_0 = (a_{01}, \ldots, a_{0H}) \in [0,1]^H$. Mathematically, we may express this PP as
\begin{align}
    \pi_{\text{PP}}(\bm{\beta}, \phi | D_0, \bm{a}_0, \pi_0) \propto \left[\prod_{h = 1}^{H} L(\bm{\beta}, \phi | D_{0h})^{a_{0h}}\right] \pi_0(\bm{\beta}, \phi),
    \label{eq:pp_fixeda0}
\end{align}
In the \textbf{hdbayes} package, we implement the following initial prior
\begin{align}
    \beta_j &\sim N(\mu_{0j},  \sigma_{0j}^2) \text{ for } j = 1, \ldots, p, \notag \\
    \phi &\sim N^{+}(\alpha_0, \gamma_0^2),
    \label{eq:pp_fixeda0_initialprior}
\end{align} 
where $N^{+}(\mu, \sigma^2)$ denotes the normal distribution with mean $\mu$ and variance $\sigma^2$ truncated from below at zero (i.e., the half-normal distribution).
Thus, the initial prior assumes mutual independence among $(\bm{\beta}, \phi)$. Of course, the full power prior assumes dependence through the historical data when at least one $a_{0h} > 0$.

The hyperparameters $\bm{\mu}_0 = (\mu_{01}, \ldots, \mu_{0p})'$, $\bm{\sigma}_0 = (\sigma_{01}, \ldots, \sigma_{0p})'$, $\alpha_0$, and $\gamma_0$ are elicited, but \textbf{hdbayes} provides non-informative defaults.
In particular, the defaults are, $\bm{\mu}_0 = \bm{0}_p$, $\bm{\sigma}_0 = 10 \cdot \bm{1}_p$, $\alpha_0 = 0$, and $\gamma_0 = 10$, where $\bm{0}_q$ denotes the $q\times 1$-dimensional vector of zeros. This results in an independent normal initial prior for the components of $\bm{\beta}$ with zero mean and 100 variance and a half-normal initial prior for $\phi$, i.e., $\pi_0(\phi) \propto \varphi(\phi | 0, 100) \cdot 1\{ \phi > 0 \}$, where $\varphi(\cdot | \mu, \sigma^2)$ denotes the normal density function with mean $\mu$ and variance $\sigma^2$, and $1\{ A \}$ is the indicator function equaling 1 if $A$ is true and $0$ otherwise. 


The PP is implemented in the \textbf{BayesPPD} package \citep{shen2022bayesppd} via the function \texttt{glm.fixed.a0}.
The \textbf{BayesPPD} package uses Gibbs sampling where feasible (e.g., the normal linear model) and slice sampling \citep{neal2003slice} otherwise.
Slice samplers are easier to tune than Metropolis-type samplers, but it is difficult to implement multivariate versions of slice samplers, so that most implementations conduct slice sampling on the full conditional distributions. Unfortunately, these samplers can be slow and yield poor mixing for high dimensional problems.

The implementation allows for binomial models with the number of trials exceeding 1, which is a limitation of the implementation in the \textbf{hdbayes} implementation \texttt{glm.pp} (although one could always de-collapse the data).
Both \textbf{BayesPPD} and \textbf{hdbayes} allow the utilization of multiple historical data sets. However, the \textbf{BayesPPD} package does not have an implementation of inverse-Gaussian or gamma distributions, and thus is not comprehensive of all GLMs.
The syntax of \texttt{glm.fixed.a0} is somewhat less user friendly for the novice R user, as it does not utilize the \texttt{formula} class to construct the response variables and design matrix nor does it use the convenient \texttt{family} class to provide the distribution and link function.
Finally, the link functions in the \textbf{BayesPPD} are not as exhaustive as those offered in the \texttt{link-glm} class (e.g., the \texttt{cauchit} link is not available). 

A second package by the same authors as \textbf{BayesPPD}, \textbf{BayesPPDSurv} \citep{BayesPPDSurv_pkg}, implements the normalized power prior in design settings for time-to-event outcomes. However, the syntax is not improved from the \textbf{BayesPPD} implementation. While the \textbf{hdbayes} package cannot directly handle time-to-event outcomes, we show in Section~\hyperref[sec:dataanalysis_tte]{4.2}
how time-to-event data can be analyzed via the Poisson likelihood representation of the proportional hazards model with piecewise constant baseline hazards, which is precisely the same time-to-event model implemented in \textbf{BayesPPDSurv}. 

\subsection{Normalized power prior}
\label{sec:npp}
The PP in (\ref{eq:pp_fixeda0}) can be sensitive to the choice of $\bm{a}_0$. As we are in general uncertain about what $\bm{a}_0$ should be, one way to overcome this sensitivity is to treat $\bm{a}_0$ as random.
When $\bm{a}_0$ is treated as random, however, a normalizing constant must be estimated, otherwise the resulting posterior disobeys the likelihood principle \citep{duan2006evaluating,neuenschwander2009note}.
This leads to the normalized power prior (NPP), which, for multiple historical data sets, is given by
\begin{align}
    \pi_{\text{NPP}}(\bm{\beta}, \phi, \bm{a}_0 | D_0, \pi_0)
    &= \prod_{h=1}^H \pi_{\text{PP}}\left(\bm{\beta}, \phi | D_{0h}, a_{0h}, \pi_{0}^{1/H}\right)
    \pi(a_{0h})
    ,\notag \\
    &= \prod_{h=1}^H
    \frac{ L(\bm{\beta}, \phi | D_{0h})^{a_{0h}}
    \pi_0(\bm{\beta}, \phi)^{1/H}}
    {Z_h(a_{0h})} \pi(a_{0h})
    ,\notag\\
    &= \left[\prod_{h = 1}^{H} 
        \frac{L(\bm{\beta}, \phi | D_{0h})^{a_{0h}}
        }{Z_h(a_{0h})}
        \pi(a_{0h})
    \right]
    \pi_0(\bm{\beta}, \phi)
    \label{eq:npp}
\end{align}
where $Z_h(a_{0h}) = \int_{\mathbb{R}^p} \int_{0}^{\infty} L(\bm{\beta}, \phi | D_{0h}) \pi_0(\bm{\beta}, \phi)^{1/H} d\phi ~d\bm{\beta}$ is a normalizing constant, $\pi(a_{0h})$ is a prior on $a_{0h}$ (implemented as a beta prior), and the rest of the notation is the same as in Section \hyperref[sec:pp]{2.2}.
In most settings, the function $Z_h(\cdot)$ is analytically intractable, and it must therefore be estimated numerically. 

The approach taken in the \textbf{hdbayes} package is a simplified version of the two-step approach described by \cite{carvalho2021normalized}.
The algorithm is summarized in Algorithm \ref{alg:npp}. Bridge sampling in Algorithm \ref{alg:npp} is conducted via the \textbf{bridgesampling} package \citep{Gronau2020}.
Because the smooth function $\hat{Z}_h$ cannot be directly implemented in Stan, a grid $\mathcal{A}^* = \{ 0 = \alpha_1^* < \alpha_2^* < \cdots < \alpha_L^* = 1 \}$ with predicted values $\hat{Z}_h(\mathcal{A}^*)$ is provided to Stan, where $L \ge T$ and linear interpolation is utilized to estimate the normalizing constant between provided values in $\hat{Z}_h(\mathcal{A}^*)$.
This makes gradient evaluation feasible, which is a requirement for HMC methods.

\begin{algorithm}
  \SetAlgoLined
  \KwData{Input historical data $D_0 = \{ D_{0h}, h = 1, \ldots, H \}$, grid of points $\bm{\alpha} = \{ 0 = \alpha_0 < \alpha_1 < \ldots, < \alpha_T = 1 \}$, number of PP MCMC samples $M_h$, number of posterior samples $M$.
  }
  \KwResult{$M$ posterior samples using the NPP in \eqref{eq:npp}.}
  
  \For{$h \leftarrow 1$ \KwTo $H$}{
    \For{$t \leftarrow 1$ \KwTo $T$}{
        Obtain $M_h$ MCMC samples of the PP in \eqref{eq:pp_fixeda0} with historical data $D_{0h}$, initial prior $\pi_0^{1/H}$, and discounting parameter $\alpha_h$.
    \\
        Conduct bridge sampling to obtain an estimate $\widehat{Z_h(\alpha_t)}$ of $Z_h(\alpha_t)$ using the $M_h$ samples from the prior.
    }
    Use LOESS to obtain smooth estimate $\hat{Z}_h(\alpha)$ of $Z_h(\alpha)$ via $\{ \widehat{Z_h(\alpha_1)}, \ldots, \widehat{Z_h(\alpha_T)} \}$
    \\
    Obtain $M$ posterior samples using the NPP by substituting $\hat{Z}_h$ for $Z_h$ in \eqref{eq:npp}
}
  \caption{Posterior of Normalized Power Prior}
  \label{alg:npp}
\end{algorithm}

The function \texttt{glm.npp.lognc} in the \textbf{hdbayes} package estimates the logarithm of the normalizing constant for a single value $\alpha_{t}$ of the discounting parameter, with syntax similar to the \texttt{glm} function in the \textbf{stats} package \citep{rlanguage}.
In particular, sampling of (\ref{eq:pp_fixeda0}) is conducted via \textbf{cmdstanr} and then the logarithm of the normalizing constant, $\log Z_h(\alpha_{t})$ is estimated via the \textbf{bridgesampling} package.
We note that one may utilize parallel computing in order to obtain the estimated normalizing constants faster by using the \textbf{parallel} package \citep{rlanguage}, which is included as a part of \texttt{R}. 

After estimating the function $Z_h(\alpha_{t})$ and conducting smoothing, posterior samples using the NPP (\ref{eq:npp}) may be obtained via the \texttt{glm.npp} function.
The syntax for this function and \texttt{glm.pp}, as described in Section \hyperref[sec:pp]{2.2}, is virtually identical, except for \texttt{glm.npp} users must supply values for the arguments \texttt{a0.lognc} (the points in the grid) and \texttt{lognc} (the smoothed estimated logarithm of the normalizing constants) obtained as described above.

The \texttt{R} package \textbf{NPP} offers an implementation of the NPP for GLMs. The \textbf{NPP} package uses independence or random-walk metropolis proposals, which are fast but can be difficult to tune, often resulting in highly correlated samples. Moreover, the \textbf{NPP} package uses a Laplace approximation \citep{tierney1986accurate} to estimate the normalizing constant, which is much faster than the implementation in \textbf{hdbayes} but can be highly inaccurate for smaller samples and/or high dimensional problems. Finally, each distribution in the exponential family corresponds to a different function in \textbf{NPP} package, making the syntax somewhat awkward.

The aforementioned \textbf{BayesPPD} package also offers an implementation of the NPP. The approach is a two-step approach similar to the algorithm above. The main difference between \textbf{BayesPPD} and \textbf{hdbayes} besides those mentioned in Section \hyperref[sec:pp]{2.2} is that \textbf{BayesPPD} conducts slice sampling to sample from the prior density (\ref{eq:pp_fixeda0}), while HMC is utilized for the \textbf{hdbayes} package. \textbf{BayesPPD} also has different syntax for each distribution, making it somewhat more cumbersome to use in practice.

It is worth noting that, under a conjugate (multivariate normal-gamma) initial prior, the normalizing constant of the PP for the normal linear model is known and does not need to be estimated.
If a user calls \texttt{glm.npp} with \texttt{family = gaussian(`identity')}, the function requires users input a grid of estimated normalizing constants.
The \textbf{hdbayes} package offers an alternative way to sample from the posterior using the NPP for the normal linear model through the \texttt{lm.npp} function, which is a one-step approach that does \emph{not} require estimation of the normalizing constant before posterior sampling.

\subsection{Normalized asymptotic power prior}
\label{sec:napp}
\citet{ibrahim2000power} showed that, under large samples of the historical data set, the PP in \eqref{eq:pp_fixeda0_singledataset} converges to a multivariate normal density, i.e.,
\begin{align}
  \pi_{\text{PP}}(\bm{\beta}, \phi | D_{0h}, a_{0h}, \pi_0) \overset{n_{0h} \to \infty}{\to} \varphi\left( \bm{\beta}, \phi \left| (\hat{\bm{\beta}}_{0h}', \hat{\phi}_{0h})', ~a_{0h}^{-1} \left[ I(\hat{\bm{\beta}}_{0h}, \hat{\phi}_{0h} | D_{0h} ) \right]^{-1} \right) \right.,
  \label{eq:app}
\end{align}
where $\varphi(\cdot | \bm{\mu}, \bm{\Sigma})$ is the multivariate normal density function with mean $\bm{\mu}$ and covariance matrix $\bm{\Sigma}$, $\hat{\bm{\beta}}_{0h}$ and $\hat{\phi}_{0h}$ are the maximum likelihood estimates (MLEs) of $(\bm{\beta}, \phi)$ under the historical data $D_{0h}$, and $I(\cdot | D_{0h})$ is the Fisher information matrix (i.e., the expectation of the negative Hessian matrix of the log posterior density) based on the GLM likelihood for historical data $D_{0h}$.
The right-hand-side of (\ref{eq:app}) has been referred to as the ``asymptotic power prior'' \citep{ibrahim2015power}

Since the right-hand-side of (\ref{eq:app}) is properly normalized, there is no need to estimate a normalizing constant if we treat $a_{0h}$ as random.
However, since the dispersion parameter $\phi$ is positive, a very large historical data set may be required for the normal approximation to work well due to skewness.
We may thus take the transformation $\tau = \log \phi$.
By the invariance property of MLEs, we have $\hat{\tau}_{0h} = \log \hat{\phi}_{0h}$, and the Jacobian matrix of this transformation is given by
$$
  J(\bm{\beta}, \tau) = \begin{pmatrix}
     \pdv{\bm{\beta}}{\bm{\beta}'} & \pdv{\bm{\beta}}{\tau} \\
     \pdv{\phi}{\bm{\beta}'}       & \pdv{\phi}{\tau}
  \end{pmatrix}
  = \begin{pmatrix}
     \bm{I}  & \bm{0}_p \\
     \bm{0}_p' & \exp{\tau}
  \end{pmatrix},
$$
so that the Fisher information for historical data $D_{0h}$ is given by $I(\bm{\beta}, \tau) = J(\bm{\beta}, \tau)' I(\bm{\beta}, \exp\{\tau\} | D_{0h}) J(\bm{\beta}, \tau)$

The package \textbf{hdbayes} offers an implementation of what we call the normalized asymptotic power prior (NAPP) using the transformation $\tau = \log \phi$. Let $\bm{\theta} = (\bm{\beta}', \tau)'$ and $\hat{\bm{\theta}}_{0h} = (\hat{\bm{\beta}}_{0h}', \hat{\tau}_{0h})'$. The NAPP is given by
$$
\pi_{\text{NAPP}}(\bm{\theta}, \bm{a}_0 | D_0) = \prod_{h=1}^H \varphi\left( \bm{\theta} \left| \hat{\bm{\theta}}_{0h}, ~a_{0h}^{-1} [I(\hat{\bm{\theta}}_{0h} | D_{0h})]^{-1}\right) \right. \pi(a_{0h}),
$$
where we take an independent Beta prior for each $a_{0h}, h = 1, \ldots, H$.
The primary advantage of the NAPP over the NPP is that there is no need to estimate a normalizing constant, so that the implementation is a one-step approach. However, when a normal distribution is a poor approximation for the likelihood function, the NAPP might be overly informative. Posterior samples utilizing the NAPP may be obtained using the \texttt{glm.napp} function, which only requires a \texttt{formula}, \texttt{family}, and a list of \texttt{data.frame}s giving the current and historical data sets.

\subsection{Bayesian hierarchical model}
\label{sec:bhm}

The Bayesian hierarchical model (BHM) is arguably the most widely used Bayesian model for informative prior elicitation.
The BHM assumes that the parameters for the current and historical data sets are different, but come from the same distribution whose hyperparameters themselves are treated as random.

Let $(\bm{\beta}', \phi)'$ denote the GLM parameters for the current data set, where $\bm{\beta} = (\beta_1, \ldots, \beta_p)'$ Let $(\bm{\beta}_{0h}', \phi_{0h})'$ denote the GLM parameters for the historical data set $D_{0h}$, $h = 1, \ldots, H$, where $\bm{\beta}_{0h} = (\beta_{0h1}, \ldots, \beta_{0hp})'$.
The BHM as implemented in \textbf{hdbayes} may be expressed hierarchically as
\begin{align}
    \mu_j | \mu_{0j}, \sigma_{0j} &\sim N(\mu_{0j}, \sigma_{0j}^2)
        , \ \ j = 1, \ldots, p
    ,\notag \\
    \sigma_j | m_j, s_j &\sim N^+(m_j, s_j^2)
        , \ \ j = 1, \ldots, p
    ,\notag \\
    \beta_j, \beta_{0hj} | \mu_j, \sigma_j &\overset{\text{i.i.d.}}{\sim} N(\mu_j, \sigma_j^2)
        , \ \ h = 1, \ldots, H
        , \ \ j = 1, \ldots, p
    ,\notag \\
    \phi | m_0, s_0 &\sim N^{+}(m_0, s_0^2)
        , \notag \\
    \phi_{0h} | m_{0h}, s_{0h} &\sim N^{+}(m_{0h}, s_{0h}^2)
        , \ \  h = 1, \ldots, H
    ,\notag \\
    y_i    | \bm{\beta}, \phi &\sim f(y_i | \bm{\beta}, \phi)
        , \ \ i = 1, \ldots, n
    , \notag \\
    y_{0hi} | \bm{\beta}_{0h}, \phi_{0h} &\sim f(y_{0hi} | \bm{\beta}_{0h}, \phi_{0h}), 
        \ \ h = 1, \ldots, H, 
        \ \ i = 1, \ldots, n_{0h}, 
    \label{eq:bhm}
\end{align}
where $f(\cdot | \bm{\beta}, \phi)$ is the density (or mass) function corresponding to the GLM likelihood in \eqref{eq:glm_likelihood}, 
$\mu_j$ is referred to as the global (or meta-analytic) mean, 
$\sigma_j$ is the global standard deviation (measuring how different the parameters are between the data sets), 
and $\bm{\xi} = ( m_0, s_0, \{(m_{0h}, s_{0h}), h = 1, \ldots, H \}, \{ (\mu_{0j}, \sigma_{0j}, m_j, s_j) , j = 1, \ldots, p \} )$ are elicited hyperparameters. The hyperparameters $s_j$ are the most crucial ones, as they must often be chosen to be somewhat subjective and reflect most of the borrowing properties under the BHM.

Let $\bm{\theta}$ denote all parameters to be sampled in \eqref{eq:bhm}. The posterior density of \eqref{eq:bhm} may be expressed as
\begin{align}
    p_{\text{BHM}}(\bm{\theta} | D, D_0, \bm{\xi}) \propto
    L(\bm{\beta}, \phi | D)
    \left[
        \prod_{h=1}^H L(\bm{\beta}_{0h}, \phi_{0h} | D_{0h})
    \right]
    \pi_{\text{BHM}}(\bm{\theta})
    ,
    \label{eq:bhm_post}
\end{align}
where
\begin{align}
    \pi_{\text{BHM}}(\bm{\theta}) &=
    \varphi^+(\phi | m_0, s_0^2)
    \left[ \prod_{h=1}^H \varphi^+(\phi_{0h} | m_{0h}, s_{0h}^2) \right] \cdot \notag \\
    &~~~~~ \left\{
    \prod_{j=1}^p 
    \left[ \varphi(\mu_j | \mu_{0j}, \sigma_{0j}^2) \varphi^+(\sigma_j | m_j, s_j^2)
    \varphi(\beta_j | \mu_j, \sigma_j^2) \prod_{h=1}^H \varphi(\beta_{0hj} | \mu_j, \sigma_j^2)
    \right]
    \right\},
    \label{eq:bhm_prior}
\end{align}
$\varphi^+(\cdot | a, b^2)$ denotes the half-normal density with location $a$ and scale parameter $b$, and $\varphi(\cdot | a, b^2)$ denotes the normal density with mean $a$ and standard deviation parameter $b$. 

In \textbf{hdbayes}, posterior inference for the BHM may be conducted using the \texttt{glm.bhm} function. The default choices for the hyperparameters are $\mu_{0j} = 0$, $\sigma_{0j} = 10$, $m_j = 0$, $s_j = 1$, $m_0 = m_{0h} = 0$, $s_0 = s_{0h} = 10$ for $j = 1, \ldots, p$ and $h = 1, \ldots, H$. The value $s_j = 1$ is somewhat informative, but designed to encourage information borrowing. In general, $s_j$ cannot be chosen to be large unless there are many ($e.g. \ge 10$) historical data sets, which is rarely the case.

\subsection{Robust meta-analytic predictive prior}
\label{sec:RMAP}

\cite{schmidli2014robust} showed that the BHM in \eqref{eq:bhm_post} induces a prior for the current data regression coefficients (referred to as the meta-analytic predictive (MAP) prior), which is given by
\begin{align}
    \pi_{\text{MAP}}(\bm{\beta} | D_0)
        &= \int \int \left[\prod_{j=1}^p \varphi(\beta_j | \mu_j, \sigma_j^2) \right] \pi(\bm{\mu}, \bm{\sigma} | D_0) d\bm{\mu} d\bm{\sigma},
        \label{eq:map}
\end{align}
where $\bm{\mu} = (\mu_1, \ldots, \mu_p)$, $\bm{\sigma} = (\sigma_1, \ldots, \sigma_p)$, and $\pi(\bm{\mu}, \bm{\sigma} | D_0)$ is the posterior density of $(\bm{\mu}, \bm{\sigma})$ obtained by a BHM using only the historical data, i.e.,
\begin{align}
    \pi(\bm{\mu}, \bm{\sigma} | D_0) &= 
    \int \int p_{\text{BHM}}(\bm{\mu}, \bm{\sigma}, \bm{\beta}_0, \bm{\phi}_0 | D_0) d\bm{\beta}_0 d\bm{\phi}_0
    ,\notag \\
    &\propto \int \int \left\{\prod_{h=1}^H L(\bm{\beta}_{0h}, \phi_{0h} | D_{0h}) \varphi^+(\phi_{0h} | m_{0h}, s_{0h}^2) \prod_{j=1}^p \varphi(\beta_{0hj} | \mu_j, \sigma_j^2)
    \right\}
    d\bm{\beta}_0 d\bm{\phi}_0,
    \label{eq:map_mean_sd}
\end{align}
where $p_{\text{BHM}}(\cdot | D_0)$ is the posterior density of the BHM for the historical data and is given in \eqref{eq:bhm_post}, $\bm{\beta}_0 = (\bm{\beta}_{01}', \ldots, \bm{\beta}_{0H}')'$, and $\bm{\phi}_0 = (\phi_{01}, \ldots, \phi_{0H})'$.

The \emph{robust} meta-analytic predictive (RMAP) prior \citep{schmidli2014robust} is a two-part mixture prior consisting of a meta-analytic predictive (MAP) prior using the historical data sets and a vague (i.e., non-informative) prior. For an arbitrary parameter vector $\bm{\theta}$, the RMAP prior may be expressed as $\pi_{\text{RMAP}}(\bm{\theta} | D_0, \gamma) = \gamma \pi_{\text{MAP}}(\bm{\theta} | D_0) + (1 - \gamma) \pi_{v}(\bm{\theta})$, where $\gamma \in [0, 1]$ is an elicited hyperparameter that controls the influence of the historical data on the prior, and $\pi_v$ is the vague prior. Note that if $\gamma = 0$, the RMAP prior is simply the vague prior. Conversely, if $\gamma = 1$, the RMAP is the MAP, which results in the same posterior density as the BHM when combined with the current data likelihood.

Although \cite{schmidli2014robust} recommends to use a finite mixture of conjugate priors to approximate the BHM, it can be difficult and time consuming to come up with an appropriate approximation. Instead, the approach taken by \textbf{hdbayes} is to use the marginal likelihood of the non-informative and meta-analytic predictive priors. Specifically, note that
\begin{align}
    p_{\text{RMAP}}(\bm{\beta}, \phi | D, D_0, \gamma) &= 
    \frac{
    L(\bm{\beta}, \phi | D) \left[ \gamma \pi_I(\bm{\beta}, \phi | D_0) + (1 - \gamma) \pi_V(\bm{\beta}, \phi) \right]
    }{
        \int \int
        L(\bm{\beta}^*, \phi^* | D) \left[ \gamma \pi_I( \bm{\beta}^* \phi^* | D_0) + (1 - \gamma) \pi_V( \bm{\beta}^*, \phi^* ) \right] 
        d\bm{\beta}^*, d\phi^*
    }
    ,
    \notag \\
    &= \tilde{\gamma} p_I(\bm{\beta}, \phi | D, D_0) + (1 - \tilde{\gamma}) p_V(\bm{\beta}, \phi | D),
\end{align}
where $p_I(\bm{\beta}, \phi | D, D_0) = L(\bm{\beta}, \phi | D) \pi_I(\bm{\beta}, \phi | D_0) / Z_I(D, D_0)$ is the posterior density under the informative prior, $p_V(\bm{\beta}, \phi | D) = L(\bm{\beta}, \phi | D) \pi_V(\bm{\beta}, \phi) / Z_V(D)$ is the posterior density under the vague prior, and
\begin{align}
    \tilde{\gamma} = \frac{ \gamma Z_I(D, D_0) }{ \gamma Z_I(D, D_0) + (1 - \gamma) Z_V(D_0) }
    \label{eq:rmap_weight}
\end{align}
is the updated mixture weight. The normalizing constants $Z_I(D, D_0)$ and $Z_V(D)$ are estimated via \textbf{bridgesampling} \citep{Gronau2020} in \textbf{hdbayes}. This approach obviates the need to come up with a finite mixture approximation to the informative prior, and we find it to be more computationally convenient. Details for computing the normalizing constants are given in Section \hyperref[sec:normconst]{3}. The Algorithm for the implementation is presented in Algorithm \ref{alg:rmap}

\begin{algorithm}
  \SetAlgoLined
  \KwData{Input current data $D$, historical data $D_0 = \{ D_{0h}, h = 1, \ldots, H \}$, number of posterior samples $M$, number of prior samples $M_0$, hyperparameters $\bm{\xi}$, maximum number of mixture components $K$.
  }
  \KwResult{$M$ posterior samples using the RMAP prior.}
  Obtain $M$ samples from a BHM using the current and historical data, $p_{I}(\bm{\theta} | D, D_0)$.
  \\ Obtain $M$ samples using the posterior under a vague prior, $p_V(\bm{\theta} | D_0)$.
  \\ Compute the normalizing constants $Z_I(D, D_0)$ and $Z_V(D)$ via \textbf{bridgesampling}. Then compute $\tilde{\gamma}$ in \eqref{eq:rmap_weight}.
  \\ For $m = 1, \ldots, M$, draw $j_m \sim \text{Bernoulli}(\tilde{\gamma})$. If $j_m = 1$, sample $\bm{\theta}_m$ from the posterior samples pertaining to $p_I(\cdot | D, D_0)$. Otherwise, sample $\bm{\theta}_m$ from the posterior samples pertaining to $p_V(\cdot | D)$.
  \caption{Posterior under Robust Meta-Analytic Predictive Prior}
  \label{alg:rmap}
\end{algorithm}

\subsection{Commensurate prior}
The commensurate prior (CP) of \cite{hobbs2012commensurate} is a hierarchical prior. In the traditional CP, the current data regression coefficients, $\bm{\beta} = (\beta_1, \ldots, \beta_p)'$ are assumed to be normally distributed with mean being the historical data regression coefficients, $\bm{\beta}_0 = (\beta_{01}, \ldots, \beta_{0p})'$, and a hierarchical precision parameter. Specifically, the CP assumes $\beta_j \sim N(\beta_{0j}, \tau_j^{-1})$, where $\tau_j$ is referred to as the ``commensurability parameter'' for regression coefficient $j$. The parameter $\tau_j$ measures how compatible the current and historical data are based on the $j^{th}$ covariate, with higher values indicating a larger degree of commensurability.

To our knowledge, a CP for multiple historical data sets has not yet been developed. In \textbf{hdbayes}, we implement the CP by assuming that the historical data sets all have the same regression coefficients (but different dispersion parameters if applicable).
Expressed hierarchically, the CP as implemented in \textbf{hdbayes} is given by
\begin{align}
   &\beta_{0j} | \mu_{0j}, \sigma_{0j} \sim N(\mu_{0j}, \sigma_{0j}^2), \ \ j = 1, \ldots, p
   , \notag \\
   &\phi | m_0, s_0 \sim N^+(m_0, s_0^2)
   ,\notag \\
   &\phi_{0h} | m_{0h}, s_{0h} \sim N^+(m_{0h}, s_{0h}^2), \ \ h = 1, \ldots, H 
   ,\notag \\
   &\beta_j | \beta_{0j}, \tau_j \sim N\left( \beta_{0j}, \tau_j^{-1} \right),
        \ \ j = 1, \ldots, p
    ,\notag \\
    &\tau_j | p_{\text{spike}}, \mu_{\text{spike}}, \sigma_{\text{spike}}, \mu_{\text{slab}}, \sigma_{\text{slab}} \sim 
    p_{\text{spike}} N^+(\mu_\text{spike}, \sigma_{\text{spike}}^2)
    + (1 - p_{\text{spike}}) N^+(\mu_{\text{slab}}, \sigma_{\text{slab}}^2)
    , \notag \\
    &y_i    | \bm{\beta}, \phi \sim f(y_i | \bm{\beta}, \phi), \ \  i = 1, \ldots, n
    , \notag \\
    &y_{0hi} | \bm{\beta}_0, \phi_{0h} \sim f(y_{0hi} | \bm{\beta}_0, \phi_{0h}), \ \ h = 1, \ldots, H, \ \  i = 1, \ldots, n_{0h}
    \label{eq:comm_hm}
\end{align}
The $\tau_j$'s are referred to as ``commensurability parameters,'' and they quantify how compatible the historical data are with the current data (higher values indicate a higher degree of compatibility). Following \cite{hobbs2012commensurate}, we elicit a spike-and-slab prior on the commensurability parameters. When there is only one historical data set (i.e., for $H = 1$), the CP implemented in \textbf{hdbayes} corresponds to the traditional commensurate prior of \cite{hobbs2012commensurate}. 

The commensurate prior is implemented in \textbf{hdbayes} via the function \texttt{glm.commensurate}.
The default hyperparameters are
\begin{align*}
    \mu_{0j} &= 0, \ \ j = 1, \ldots, p
    ,\\
    \sigma_{0j} &= 10, \ \ j = 1, \ldots, p
    ,\\
    m_0 &= m_{0h} = 0, \ \ h = 1, \ldots, H
    , \\
    s_0 &= s_{0h} = 10, \ \ h = 1, \ldots, H
    , \\
    p_{\text{spike}} &= 0.1
    ,\\
     \mu_{\text{spike}} &= 200
    ,\\
    \sigma_{\text{spike}} &= 0.1
    , \\
    \mu_{\text{slab}} &= 0
    , \\
    \sigma_{\text{slab}} &= 5.
\end{align*}
The default hyperparameters for the ``spike'' approximates a point mass at $\tau_j = 200$ and encourages a high degree of borrowing. Conversely, the default hyperparameters for the slab results in a density approximately uniform over $[0, 3]$ and decaying thereafter, encouraging a small amount of borrowing. In general, elicitation of the hyperparameters for the half-normal prior on the $\tau_j$'s is problem specific.

The \texttt{R} packages \textbf{psborrow} \citep{psborrow_pkg} and \textbf{psborrow2} \citep{psborrow2_pkg} conduct propensity score-based implementations of the CP, with the former relying on \textbf{rjags} to conduct MCMC sampling and the latter relying on \textbf{cmdstanr} (like \textbf{hdbayes}). 
The \textbf{psborrow2} package offers more flexibility for initial prior elicitation than does \textbf{hdbayes}.
However, initial priors are usually taken to be non-informative, so the class of distributions will not have an impact on the analysis results in the vast majority of applications.
Moreover, this increased flexibility comes at the cost of having somewhat more complicated syntax, while that in \textbf{hdbayes} is similar to the \texttt{glm} function in the \textbf{stats} package.

\subsection{Latent exchangeability prior}\label{sec:LEAP}
The latent exchangeability prior (LEAP), developed by \citet{alt2023leap}, assumes that the historical data are generated from a finite mixture model consisting of $K \ge 2$ components, with the current data generated from one component of this mixture (the first component, without loss of generality). For single historical data set settings, the posterior under the LEAP may be expressed hierarchically as
\begin{align}
    \bm{\gamma} &\sim \text{Dirichlet}(\bm{\alpha}_0)
    ,\notag \\
    \bm{\beta}, \bm{\beta}_{0k} 
        &\overset{\text{i.i.d.}}{\sim}
        N(\bm{\mu}_0, \bm{\Sigma}_0)
    , \ \ k = 2, \ldots, K
    ,\notag \\
    \phi, \phi_{0k}
        &\overset{\text{i.i.d.}}{\sim}
        N^+(m_0, s_0^2)
    , \ \ k = 2, \ldots, K
    ,\notag \\
    y_i | \bm{\beta}, \phi &\sim f(\cdot | \bm{\beta}, \phi)
    , \ \  i = 1, \ldots, n
    , \notag \\
    y_{0i} | \bm{\beta}, \bm{\beta}_0, \phi, \bm{\phi}_0, \bm{\gamma} &\sim \gamma_1 f(\cdot | \bm{\beta}, \phi) + \sum_{k=2}^K \gamma_k f(\cdot | \bm{\beta}_{0k}, \phi_{0k}), 
    , \ \ i = 1, \ldots, n_0,
    \label{eq:leap}
\end{align}
where $\bm{\gamma} = (\gamma_1, \ldots, \gamma_K)$ are the mixing probabilities, $\bm{\alpha}_0 = (\alpha_{01}, \ldots, \alpha_{0K})'$ is a concentration hyperparameter, $\bm{\mu}_0$ and $\bm{\Sigma}_0$ respectively are prior mean and covariance matrices for the $K$ regression coefficients, and $m_0$ and $s_0$ are respectively the location and scale parameter for the half-normal prior on the $K$ dispersion parameters. The default hyperparameters in \textbf{hdbayes} are $K = 2$, $\bm{\alpha}_0 = \bm{1}_K$, $\bm{\mu}_0 = \bm{0}_p$, $\bm{\Sigma}_0 = \text{diag}\{ \sigma_{0j}^2, j = 1, \ldots p \}$ with $\sigma_{0j} = 10$, all of which correspond to non-informative initial priors.

Unlike the above priors, which conduct blanket discounting of the historical data, the LEAP conducts discounting at the individual level of the historical data. Since the LEAP has not been developed for multiple historical data sets, the approach taken by \textbf{hdbayes} is to stack all $H$ historical data sets into one historical data set $D_0$ with $n_0 = \sum_{h=1}^H n_{0h}$ observations. The LEAP is implemented in the \texttt{glm.leap} function of \textbf{hdbayes}. The \textbf{hdbayes} package contains the first implementation of the LEAP in a publicly available \texttt{R} package.

\section{Model selection via marginal likelihoods}\label{sec:normconst}
Canonical Bayesian model selection proceeds by calculating the marginal likelihood (also referred to as the ``evidence''). For example, suppose the model space $\mathcal{M} = \{M_1, M_2\}$ consists of two candidate models. \cite{kass1995bayes} shows that the preference of $M_2$ over $M_1$ can be expressed as a function of the ``Bayes factor'', given by
\begin{align}
    \text{BF}(D) = \frac{Z_2(D)}{Z_1(D)} = \frac{ \int L_2(\bm{\theta}_2 | D) \pi_2(\bm{\theta}_2) d\bm{\theta}_2 }{ \int L_1(\bm{\theta}_1 | D) \pi_1(\bm{\theta}_1) d\bm{\theta}_1 },
\end{align}
where $L_j(\cdot | D)$ is the likelihood corresponding to model $M_j$ with parameters $\bm{\theta}_j$ and prior $\pi_j$ and $Z_j(D)$ is the normalizing constant of the posterior for model $M_j$ and prior $\pi_j$ (i.e., the marginal likelihood for model $M_j$). 

In \textbf{hdbayes}, many of the implemented priors are not normalized. While this does not present an issue for accurate MCMC sampling, it will in general result in marginal likelihoods that are inaccurate to a constant of proportionality. When comparing different models with different priors, regression coefficients, etc., this can result in incorrect Bayes factors. Let $\pi(\bm{\theta}) = \tilde{\pi}(\bm{\theta}) / C$ denote a prior for $\bm{\theta}$, where $\tilde{\pi}$ is the unnormalized prior and $C = \int \tilde{\pi}(\bm{\theta}) d\bm{\theta}$ is the normalizing constant for the prior. The marginal likelihood may be computed as
\begin{align}
    Z(D) &= \int L(\bm{\theta} | D) \pi(\bm{\theta}) d\bm{\theta}
    = \frac{ \int L(\bm{\theta} | D) \tilde{\pi}(\bm{\theta}) d\bm{\theta} }{ \int \tilde{\pi}(\bm{\theta}^*) d\bm{\theta}^* }
    = \frac{\tilde{Z}(D)}{C},
    \label{eq:normconst}
\end{align}
where $\tilde{Z}(D) = \int L(\bm{\theta} | D) \tilde{\pi}(\bm{\theta}) d\bm{\theta}$. 

It follows that the marginal likelihood can be computed in a two-step process. First, the normalizing constant of the unnormalized prior is estimated (e.g., via bridge sampling after taking MCMC samples of the prior). Second, the MCMC samples can be taken from the posterior using $\tilde{\pi}$ and estimating the normalizing constant of the posterior (e.g., via bridge sampling). 

We give an example of these techniques in Appendix \hyperref[sec:app_link_selection]{A}, where we illustrate how to select between the logit and probit link functions in the HIV data set (see Section~\hyperref[sec:dataanalysis_logistic]{4.1}) under the power prior using the marginal likelihood functionality in \textbf{hdbayes}. 

\subsection{Example: LEAP}
To illustrate our approach, consider the LEAP. Let $L(\bm{\theta}_1 | D)$ denote the likelihood for the current data and let $L_0(\bm{\theta}, \bm{\gamma} | D_0)$ denote the likelihood pertaining to the mixture model for the historical data, where $\bm{\theta} = (\bm{\theta}_1', \ldots, \bm{\theta}_K')'$ and $\bm{\gamma} = (\gamma_1, \ldots, \gamma_K)'$. Also, let $\pi_0(\bm{\theta}, \bm{\gamma})$ denote the initial prior. The marginal likelihood may be expressed as
\begin{align}
    Z(D) &= \int L(\bm{\theta}_1 | D) \pi_\text{LEAP}(\bm{\theta}_1 | D_0) d\bm{\theta}_1
    %
    = \int \int L(\bm{\theta}_1 | D) \frac{ L_0(\bm{\theta}, \bm{\gamma} | D_0)) \pi_0(\bm{\theta}, \bm{\gamma})  }{ \int \int L_0(\bm{\theta}^*, \bm{\gamma}^* | D_0) d\bm{\theta}^* d\bm{\gamma}^* } d\bm{\theta} d\bm{\gamma}
    ,\notag \\ &
    = \frac{ \int \int L(\bm{\theta}_1 | D) L_0(\bm{\theta}, \bm{\gamma} | D_0) \pi_0(\bm{\theta}, \bm{\gamma}) d\bm{\theta} d\bm{\gamma} }
    {\int \int L_0(\bm{\theta}^*, \bm{\gamma}^* | D_0) d\bm{\theta}^* d\bm{\gamma}^*},
    \label{eq:normconst_leap}
\end{align}
which takes the same form as \eqref{eq:normconst}. To estimate the marginal likelihood of the LEAP in \textbf{hdbayes}, we first estimate the marginal likelihood of the Bayesian mixture model via bridgesampling to provide an estimate of the denominator of the last equality in \eqref{eq:normconst_leap}. Second, we fit the LEAP and compute the normalizing constant via \textbf{bridgesampling}, providing an estimate for the numerator of the last equality in \eqref{eq:normconst_leap}. Taking the ratio provides an estimate of $Z(D)$ for the LEAP. 

\section{Data analyses}
\label{sec:dataanalysis}
\subsection{Logistic regression example}
\label{sec:dataanalysis_logistic}

We now illustrate the functionality in \textbf{hdbayes} by analyzing the data on two clinical trials presented in Section 4.2 of \citet{chen1999prior} using logistic regression to model progression to HIV in a cohort of patients treated with zidovudine (AZT). 
More details and full output can be found in the vignette ``AIDS-Progression'' that accompanies \textbf{hdbayes}.
The historical data comes from the ACTG019 \citep{Volberding1990} study, which
was a double-blind placebo-controlled clinical trial comparing zidovudine (AZT)
with a placebo in people with CD4 cell counts of less than 500.
The sample size of complete observations for this study was $n_0 = 822$.
The response variable (\texttt{outcome}) for these data is binary, with 1 indicating death,
development of AIDS or AIDS-related complex (ARC) and 0 otherwise.
We will use the following covariates: CD4 cell count (\texttt{cd4}), \texttt{age}, \texttt{treatment} and \texttt{race}.

Interest lies in analyzing the data from a more recent study known as ACTG036 \citep{Merigan1991}, for which $n=183$ observations are available. 
We will use the methods in \textbf{hdbayes} to construct informative priors using the ACTG019 data as historical data in order to analyze the data from ACTG036 study, for which we will use the same four covariates.
To facilitate computation, we will center and scale the continuous covariates (\texttt{age} and \texttt{cd4}).
In general, we recommend this centering procedure in order to keep coefficients on roughly the same scale and thus avoid difficult posterior geometries for the Stan dynamic Hamiltonian Monte Carlo (dHMC) procedure to explore.
The data preparation can be accomplished by calling
\begin{verbatim}
library(hdbayes)
library(posterior)
library(dplyr)
library(parallel)

data(actg019)
data(actg036)
age_stats <- with(actg036,
                  c('mean' = mean(age), 'sd' = sd(age)))
cd4_stats <- with(actg036,
                  c('mean' = mean(cd4), 'sd' = sd(cd4)))
actg036$age <- ( actg036$age - age_stats['mean'] ) / age_stats['sd']
actg019$age <- ( actg019$age - age_stats['mean'] ) / age_stats['sd']
actg036$cd4 <- ( actg036$cd4 - cd4_stats['mean'] ) / cd4_stats['sd']
actg019$cd4 <- ( actg019$cd4 - cd4_stats['mean'] ) / cd4_stats['sd']
\end{verbatim}
Now, we set up the analysis by creating a formula and (GLM) family, as well as obtaining the maximum likelihood estimates (MLEs) of the regression coefficients in both current and historical data models using \texttt{stats::glm()}. We also specify a list of data sets, with the first element being the current data and the second element being the historical data. In \textbf{hdbayes}, we assume that the first specified data set in the list is the current data sets, and any other data sets are historical.
\begin{verbatim}
formula <- outcome ~  age + race + treatment + cd4
p       <- length(attr(terms(formula), "term.labels")) ## number of predictors
family  <- binomial('logit')

fit.mle.cur  <- glm(formula, family, actg036)
fit.mle.hist <- glm(formula, family, actg019)

the.data <- list(actg036, actg019)   
\end{verbatim}
We can now briefly inspect the confidence intervals for the coefficients in the historical data (ACTG019) model:
\begin{verbatim}
round(confint(fit.mle.hist), 3)

             2.5 % 97.5 %
(Intercept) -6.901 -2.440
age          0.114  0.870
race        -0.026  4.424
treatment   -1.357 -0.158
cd4         -1.031 -0.396
\end{verbatim}
and for the coefficients in the current data (ACTG036) model:
\begin{verbatim}
round(confint(fit.mle.cur), 3)

             2.5 % 97.5 %
(Intercept) -7.237 -1.840
age         -0.513  0.813
race        -1.932  3.141
treatment   -1.575  1.317
cd4         -2.913 -0.926
\end{verbatim}
From these confidence intervals, we can see that there is (i) substantial uncertainty in the estimates from the current data model and (ii) disagreement between the estimates of certain coefficients, in particular the one for \texttt{treatment}, which is of primary interest here. 
This motivates the use of informative priors that, whilst drawing on historical data, can accommodate prior-data conflict by allowing for discounting.

Next, we set up the computational specifications of our analysis, in which we will run four chains each with {2,500} posterior samples after a burn-in period of {1,000} in parallel (resulting in {10,000} post burn-in posterior samples):

\begin{verbatim}
ncores        <- 4
nchains       <- 4 ## number of Markov chains
iter_warmup   <- 1000 ## warmup per chain for MCMC sampling
iter_sampling <- 2500 ## number of samples post warmup per chain
\end{verbatim}
We are now prepared to fit a BHM (Section~\hyperref[sec:bhm]{2.5}), which can be achieved by simply calling its dedicated function, \texttt{glm.bhm}:

\begin{verbatim}
fit.bhm <- glm.bhm(
  formula, family, the.data,
  meta.mean.mean = 0, meta.mean.sd = 10,
  meta.sd.mean = 0, meta.sd.sd = 0.5,
  iter_warmup = iter_warmup, iter_sampling = iter_sampling,
  chains = nchains, parallel_chains = ncores,
  refresh = 0
)
\end{verbatim}
The output from \texttt{glm.bhm} is a \texttt{draws\_df} object derived from the \textbf{posterior} package. One may utilize functions in the \textbf{posterior} package \citep{posterior_pkg}, like \texttt{summarise\_draws()}, to acquire posterior inference and MCMC diagnostics. For instance, one can obtain posterior mean, standard deviation, and quantiles for both current and historical data regression coefficients as follows:

\begin{verbatim}
## function to pull out the summaries in a convenient form
## function to pull out the posterior summaries in a convenient form
get_summaries <- function(fit, pars.interest, digits = 3) {
  suppressWarnings(
    fit %>%
      select(all_of(pars.interest)) %>%
      summarise_draws(mean, sd, ~quantile(.x, probs = c(0.025, 0.5, 0.975)),
                      .num_args = list(digits = digits))
  )
}
## names of current and historical regression: coefficients
base.pars      <- c("(Intercept)", "age", "race", "treatment", "cd4")
base.pars.hist <- paste(base.pars, "hist", "1", sep="_")
get_summaries(fit = fit.bhm, pars.interest = c(base.pars, base.pars.hist))

# A tibble: 10 × 6
   variable             mean    sd `2.5%`  `50%` `97.5%`
   <chr>               <dbl> <dbl>  <dbl>  <dbl>   <dbl>
 1 (Intercept)        -4.130 0.906 -6.151 -4.049  -2.601
 2 age                 0.262 0.261 -0.307  0.283   0.729
 3 race                1.012 0.903 -0.593  0.958   2.956
 4 treatment          -0.644 0.430 -1.496 -0.648   0.231
 5 cd4                -1.247 0.369 -2.066 -1.214  -0.645
 6 (Intercept)_hist_1 -3.917 0.828 -5.791 -3.841  -2.527
 7 age_hist_1          0.456 0.185  0.097  0.455   0.822
 8 race_hist_1         1.383 0.826  0.006  1.309   3.267
 9 treatment_hist_1   -0.706 0.291 -1.287 -0.700  -0.139
10 cd4_hist_1         -0.776 0.164 -1.098 -0.777  -0.450
\end{verbatim}
Having used the BHM as an example of the sort of output one may extract from the functions available in \textbf{hdbayes}, we will now quickly describe how to fit the other models discussed in Section~\hyperref[sec:priorelicit]{2} in order to later compare their results.
To fit the commensurate prior model, a simple call to \texttt{glm.commensurate()} will suffice. The posterior inference for the current and historical data regression coefficients can be obtained similarly as in the BHM example.

\begin{verbatim}
fit.commensurate <- glm.commensurate(
  formula = formula, family = family, data.list = the.data,
  p.spike = 0.1, spike.mean = 200, spike.sd = 0.1,
  slab.mean = 0, slab.sd = 5,
  iter_warmup = iter_warmup, iter_sampling = iter_sampling,
  chains = nchains, parallel_chains = ncores,
  refresh = 0
)
base.pars.hist <- paste(base.pars, "hist", sep="_")
get_summaries(fit = fit.commensurate, 
              pars.interest = c(base.pars, base.pars.hist))

# A tibble: 10 × 6
   variable           mean    sd `2.5%`  `50%` `97.5%`
   <chr>             <dbl> <dbl>  <dbl>  <dbl>   <dbl>
 1 (Intercept)      -4.197 0.902 -6.164 -4.121  -2.643
 2 age               0.223 0.276 -0.332  0.224   0.748
 3 race              0.980 0.892 -0.585  0.917   2.917
 4 treatment        -0.605 0.472 -1.542 -0.602   0.325
 5 cd4              -1.326 0.345 -2.046 -1.308  -0.704
 6 (Intercept)_hist -3.944 0.827 -5.783 -3.855  -2.553
 7 age_hist          0.476 0.185  0.113  0.476   0.841
 8 race_hist         1.403 0.822  0.031  1.324   3.235
 9 treatment_hist   -0.706 0.294 -1.291 -0.702  -0.137
10 cd4_hist         -0.765 0.162 -1.084 -0.765  -0.445
\end{verbatim}
Notice that here we specify a spike-and-slab prior on the commensurability parameters ($\tau_j$'s).

Fitting the robust meta-analytic predictive prior (RMAP) involves setting the hyperparameter $\gamma$ (denoted as \texttt{w} in the function \texttt{glm.rmap}) which controls the amount of borrowing of historical data. The default value in \textbf{hdbayes} is $\gamma=0.1$. To sample from the posterior density under the RMAP prior, we implement Algorithm \ref{alg:rmap} in Section~\hyperref[sec:RMAP]{2.6} as follows:

\begin{verbatim}
## fit RMAP
res.rmap <- glm.rmap(
  formula = formula, family = family, data.list = the.data,
  w = 0.1,
  iter_warmup = iter_warmup, iter_sampling = iter_sampling,
  chains = nchains, parallel_chains = ncores,
  refresh = 0
)
fit.rmap <- res.rmap[["post.samples"]]
get_summaries(fit.rmap, pars.interest = base.pars)

# A tibble: 5 × 6
  variable      mean    sd `2.5%`  `50%` `97.5%`
  <chr>        <dbl> <dbl>  <dbl>  <dbl>   <dbl>
1 (Intercept) -4.283 0.995 -6.517 -4.208  -2.595
2 age          0.236 0.286 -0.355  0.255   0.746
3 race         0.847 0.979 -1.003  0.802   2.936
4 treatment   -0.540 0.527 -1.556 -0.556   0.566
5 cd4         -1.472 0.438 -2.420 -1.445  -0.724
\end{verbatim}
While the argument \texttt{w} could be omitted, we write it explicitly and encourage users to test different values.
A similar situation occurs when using the power prior model discussed in Section~\hyperref[sec:pp]{2.2}, but here we need to set the discounting parameter $a_{01}$, for which there is no default value in \textbf{hdbayes}.
In the present analysis, we elect to set $a_{01} = \frac{1}{2}\frac{n}{n_0} \approx 0.11$. The rationale behind this approach is that when historical and current data have similar sample sizes, we set $a_{01} = 0.5$. However, when the current data has much smaller sample size than the historical data set, we discourage borrowing so as to avoid overwhelming the information contained in the current data. 

\begin{verbatim}
n0      <- nrow(actg019)
n       <- nrow(actg036)
a0.star <- (n/n0) * 1/2
fit.pp  <- glm.pp(
  formula = formula, family = family, data.list = the.data,
  a0 = a0.star, ## discounting parameter
  iter_warmup = iter_warmup, iter_sampling = iter_sampling,
  chains = nchains, parallel_chains = ncores,
  refresh = 0
)
get_summaries(fit.pp, pars.interest = base.pars)

# A tibble: 5 × 6
  variable      mean    sd `2.5%`  `50%` `97.5%`
  <chr>        <dbl> <dbl>  <dbl>  <dbl>   <dbl>
1 (Intercept) -3.875 1.240 -6.745 -3.721  -1.884
2 age          0.249 0.281 -0.303  0.249   0.793
3 race         0.718 1.223 -1.258  0.578   3.509
4 treatment   -0.482 0.564 -1.624 -0.471   0.593
5 cd4         -1.280 0.326 -1.945 -1.275  -0.672
\end{verbatim}

\begin{figure}[ht]
\centering
\includegraphics[scale=.45]{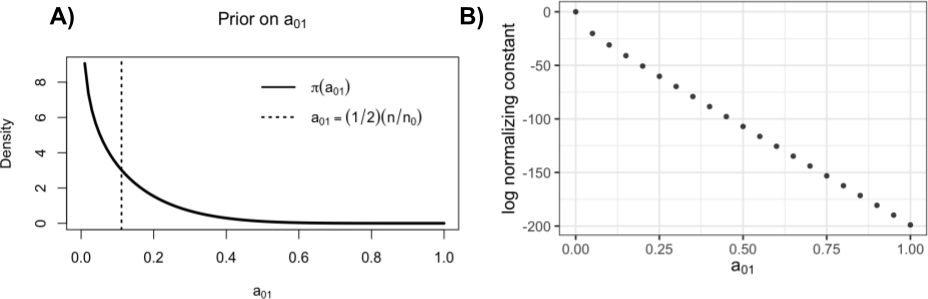}
\caption{\label{fig:a0_prior_and_normconst}
\textbf{Beta prior distribution for $a_0$ and estimated log-normalizing constants as a function of $a_{01}$ for the AIDS progression logistic regression model.}
In panel A), the vertical dotted line marks the value $1/2 (n/n_0)$.
In panel B) we show the log-normalizing constant estimated using Algorithm~\ref{alg:npp}.
Note that $\log Z_h(0) = 0$ for all $h$.
}
\end{figure}

For the normalized power prior (NPP, Section~\hyperref[sec:npp]{2.3}), it is necessary to elicit a prior distribution for $a_{01}$. 
We follow the same rationale by eliciting a beta prior with parameters $\alpha$ and $\beta$ and  mean $\alpha/(\alpha + \beta) = (1/2)(n/n_0)$ and coefficient of variation $1$.
This gives $\alpha \approx 0.77$ and $\beta \approx 6.21$.
A sketch of this prior distribution is given in Figure~\ref{fig:a0_prior_and_normconst}A.

As discussed in Section~\hyperref[sec:npp]{2.3}, for the NPP it is also necessary to estimate the normalizing constant $Z_1(a_{01})$ for a few values of $a_{01}$ so we can approximately sample from the joint posterior of $a_{01}$ and $\boldsymbol{\theta}$, the parameters of interest.
Fortunately, this task can be accomplished in parallel using the \textbf{parallel} package:
\begin{verbatim}
a0       <- seq(0, 1, length.out = 21)
histdata <- the.data[[2]]
## wrapper to obtain log normalizing constant in parallel package
logncfun <- function(a0, ...){
  hdbayes::glm.npp.lognc(
    formula = formula, family = family, histdata = histdata, a0 = a0, ...
  )
}
cl <- makeCluster(10)
clusterSetRNGStream(cl, 123)
clusterExport(cl, varlist = c('formula', 'family', 'histdata'))
## call created function
a0.lognc <- parLapply(
  cl = cl, X = a0, fun = logncfun, iter_warmup = 2*iter_warmup,
  iter_sampling = 2*iter_sampling, chains = nchains, refresh = 0
)
stopCluster(cl)
\end{verbatim}
where we were careful to specify a longer warm up period to deal with low values of $a_{01}$ which lead to tougher distributions to sample from.
The result is a \texttt{data.frame} which stores not only the values of $a_{01}$ and the corresponding estimated logarithm of the normalizing constants, but also the MCMC diagnostics so users can easily check if the results can be trusted:

\begin{verbatim}
a0.lognc <- data.frame( do.call(rbind, a0.lognc) )
head(a0.lognc) %>%
    mutate(across(where(is.numeric), round, 3))

    a0   lognc min_ess_bulk max_Rhat
1 0.00   0.000    10492.866    1.001
2 0.05 -19.658     7016.144    1.001
3 0.10 -30.372     6495.816    1.001
4 0.15 -40.382     5753.908    1.001
5 0.20 -50.115     6753.503    1.001
6 0.25 -59.686     6865.135    1.001

\end{verbatim}
Figure~\ref{fig:a0_prior_and_normconst}B displays the estimated logarithm of the normalizing constants evaluated at a grid of 21 equally-spaced values of $a_{0h} \in [0, 1]$.

We are now prepared to fit the NPP using the computations stored in \texttt{a0.lognc}:
\begin{verbatim}
fit.npp <- glm.npp(
  formula = formula, family = family, data.list = the.data,
  a0.lognc = a0.lognc$a0,
  lognc = matrix(a0.lognc$lognc, ncol = 1),
  a0.shape1 = beta.pars$a, a0.shape2 = beta.pars$b, ## beta prior on a_{01}
  iter_warmup = iter_warmup, iter_sampling = iter_sampling,
  chains = nchains, parallel_chains = ncores,
  refresh = 0
)
get_summaries(fit = fit.npp, pars.interest = c(base.pars, "a0_hist_1"))

# A tibble: 6 × 6
  variable      mean    sd `2.5%`  `50%` `97.5%`
  <chr>        <dbl> <dbl>  <dbl>  <dbl>   <dbl>
1 (Intercept) -3.836 1.179 -6.546 -3.709  -1.932
2 age          0.278 0.265 -0.251  0.279   0.781
3 race         0.802 1.170 -1.166  0.690   3.437
4 treatment   -0.539 0.534 -1.594 -0.546   0.529
5 cd4         -1.193 0.339 -1.944 -1.158  -0.615
6 a0_hist_1    0.187 0.115  0.050  0.158   0.478  
\end{example}
The normalized asymptotic power prior (NAPP) can fitted using very similar syntax:
\begin{example}
fit.napp <- glm.napp(
  formula = formula, family = family, data.list = the.data,
  a0.shape1 = beta.pars$a, a0.shape2 = beta.pars$b,
  iter_warmup = iter_warmup, iter_sampling = iter_sampling,
  chains = nchains, parallel_chains = ncores,
  refresh = 0
)
get_summaries(fit = fit.napp, pars.interest = c(base.pars, "a0_hist_1"))

# A tibble: 6 × 6
  variable      mean    sd `2.5%`  `50%` `97.5%`
  <chr>        <dbl> <dbl>  <dbl>  <dbl>   <dbl>
1 (Intercept) -3.644 1.173 -6.204 -3.541  -1.645
2 age          0.302 0.252 -0.202  0.306   0.780
3 race         0.729 1.167 -1.288  0.632   3.234
4 treatment   -0.543 0.503 -1.530 -0.547   0.468
5 cd4         -1.113 0.322 -1.840 -1.081  -0.568
6 a0_hist_1    0.188 0.128  0.026  0.158   0.502
\end{verbatim}
Note that no reference to \texttt{a0.lognc} or \texttt{lognc} is necessary; the normalizing constant is known in closed-form (see Section~\hyperref[sec:napp]{2.4}). 

As described in Section~\hyperref[sec:LEAP]{2.8}, fitting the latent exchangeability prior (LEAP) involves setting the hyperparameters $K$ and $\bm{\alpha}_0$, which are denoted as \texttt{K} and \texttt{prob.conc} in function \texttt{glm.leap}, respectively. We may sample from the posterior under the LEAP as follows:
\begin{verbatim}
fit.leap <- glm.leap(
  formula = formula, family = family, data.list = the.data,
  K = 2, prob.conc = rep(1, 2),
  iter_warmup = iter_warmup, iter_sampling = iter_sampling,
  chains = nchains, parallel_chains = ncores,
  refresh = 0
)
get_summaries(fit = fit.leap, pars.interest = c(base.pars, "gamma"))

# A tibble: 6 × 6
  variable      mean    sd `2.5%`  `50%` `97.5%`
  <chr>        <dbl> <dbl>  <dbl>  <dbl>   <dbl>
1 (Intercept) -4.206 1.025 -6.432 -4.083  -2.643 
2 age          0.300 0.183 -0.077  0.305   0.645 
3 race         1.154 1.021 -0.415  1.038   3.434
4 treatment   -0.668 0.425 -1.507 -0.680   0.228
5 cd4         -0.941 0.256 -1.507 -0.916  -0.514 
6 gamma        0.948 0.059  0.763  0.966  0.997

\end{verbatim}
In the output above, \texttt{gamma} denotes the the probability that an individual in the historical data set is exchangeable with the current data, which corresponds to $\gamma_1$ in \eqref{eq:leap}. The posterior of \texttt{gamma} suggests a high degree of exchangeability between the current and historical data sets.

\begin{figure}[!ht]
\centering
\includegraphics[width=\linewidth]{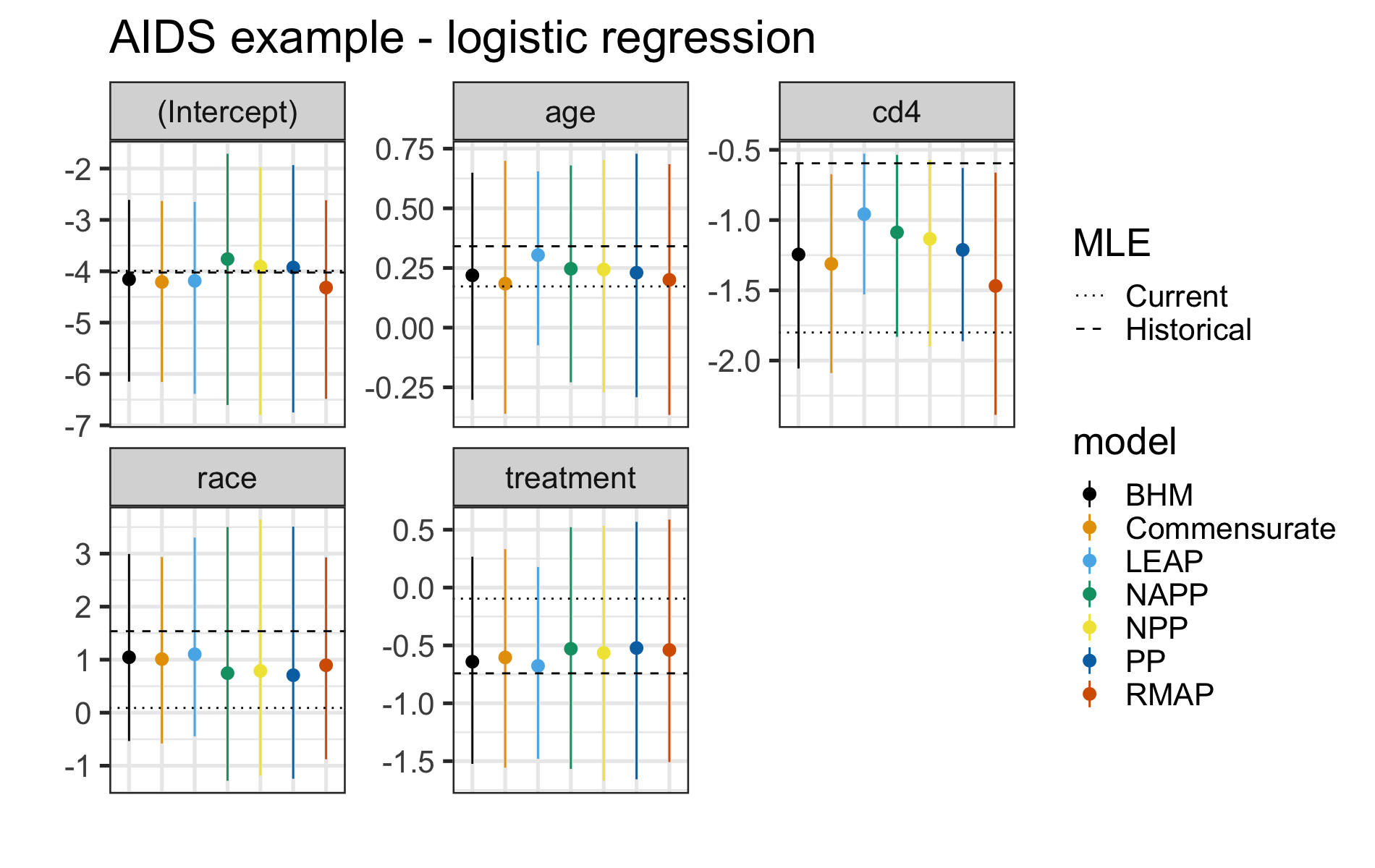}
\caption{\label{fig:models} \textbf{Coefficient estimates for the AIDS progression logistic regression model}.
For each coefficient (vertical tiles) we show the posterior mean (dot) and 95\% equal-tailed credible interval (solid vertical lines) according to each of the models used here.
Horizontal lines (dotted: ACTG036, dashed: ACTG019) mark the MLE for each coefficient.}
\end{figure}

We are now ready to compare all of the methods investigated here. The graphical comparison is shown in Figure~\ref{fig:models}.
By Figure \ref{fig:models}, there seems to be quite some variation in estimates across different methods. For instance, the BHM and RMAP more strongly pull the estimated treatment effect towards the historical MLE than the NAPP and PP do; however, the opposite behaviour occurs for the estimated coefficient for \texttt{cd4}.
For all covariates, the coefficients estimated under the LEAP are the closest to the historical MLEs. We note, however, that we did not limit the amount of borrowing for the LEAP, which can be done in principle by imposing a truncated prior on $\gamma_1$ in \eqref{eq:leap}. While some methods like the power prior lead to more uncertain estimates of the treatment effect, the BHM and LEAP lead to smaller uncertainties.
Importantly, all credible intervals include zero, indicating that despite incorporating information from a historical study with a significant treatment effect, the analysis of the current data suggests that the treatment effect could be null.

\subsection{Time-to-event example}
\label{sec:dataanalysis_tte}
Because \textbf{hdbayes} focuses on the class of GLMs, it cannot accommodate time-to-event analysis off the shelf.
However, semi-parametric time-to-event analysis may be conducted using the so-called ``Poisson-trick'' of the proportional hazards model with piecewise constant baseline hazards \citep{friedman1982piecewise}.

Let $h(t | \bm{x}) = h_0(t) e^{\bm{x}'\bm{\beta}}$ denote the conditional hazard at time $t$ based on covariates $\bm{X} = \bm{x}$ and baseline hazard $h_0(\cdot)$. For piecewise constant baseline hazards, we discretize the time axis into $J$ disjoint intervals, say, $\{ I_j = (s_{j-1}, s_j], j = 1, \ldots, J \}$ where $0 = s_0 < s_1 < \cdots < s_{J-1} < s_J = \infty$. The conditional hazard function is then given by
\begin{align}
    h(t | \bm{x}) = e^{\bm{x}'\bm{\beta}} \prod_{j=1}^J \lambda_j^{1\{t \in I_j\}},
    \label{eq:pwe_haz}
\end{align}
where $\bm{\lambda} = (\lambda_1, \ldots, \lambda_J)'$ is a vector of baseline hazards, $\bm{\beta}$ is a vector of log hazard ratios, and $1\{ A \}$ is the indicator function equaling 1 if $A$ is true and $0$ otherwise. Note that the baseline hazard in \eqref{eq:pwe_haz} is constant within an interval.

Let $R_j(t) = \left[ \min\{ t, s_j \} - s_{j-1} \right]_+$ denote the time at risk in interval $j$ at time $t$, where $[x]_+ = x \cdot 1\{x > 0\}$ is the positive part of $x$. The survival function based on \eqref{eq:pwe_haz} is given by
\begin{align}
    S(t | \bm{x}) = \exp\left\{ -\int_0^t h(s | \bm{x}) ds \right\}
    = \exp\left\{ - e^{\bm{x}'\bm{\beta}} \int_0^t h_0(s) ds \right\}
    = \prod_{j=1}^J \exp\left\{ - e^{\bm{x}'\bm{\beta}} \lambda_j R_j(t) \right\}.
\end{align}

Let $D = \{ (y_i, \nu_i, \bm{x}_i), i = 1, \ldots, n \}$ denote the current data, where, for subject $i$, $y_i = \min\{ t_i, c_i \}$ is the observed event time based on failure time $t_i > 0$ and censoring time $c_i > 0$, $\nu_i = 1\{t_i \le c_i\}$ is the event indicator, and $\bm{x}_i$ is a vector of baseline covariates. 
The current data likelihood is given by
\begin{align}
    L(\bm{\lambda}, \bm{\beta} | D) &\propto 
    \prod_{i=1}^n h(y_i | \bm{x}_i)^{\nu_i} S(y_i | \bm{x}_i)
    \propto \prod_{i=1}^n \prod_{j=1}^J
        \left[ e^{\bm{x}_i'\bm{\beta} \cdot 1\{y_i \in I_j\}} \lambda_j^{1\{y_i \in I_j\}} \right]^{\nu_i} \exp\left\{ -\lambda_j e^{\bm{x}_i'\bm{\beta}} R_j(y_i) \right\}
        ,\notag \\
        &\propto
        \prod_{i=1}^n \prod_{j=1}^J \left( \lambda_j e^{\bm{x}_i'\bm{\beta}} \right)^{\delta_{ij}}
        \exp\left\{ - \lambda_j e^{\bm{x}_i'\bm{\beta}} R_j(y_i) \right\}
        ,\notag \\
        &\propto
        \prod_{i=1}^n \prod_{j=1}^{J_i} \left. f_{\text{Pois}}\left( \delta_{ij} \right| \exp\left\{ \log \lambda_j + \bm{x}_i'\bm{\beta} + \log R_j(y_i) \right\} \right),
        \label{eq:pweph_like}
\end{align}
where $J_i$ represents the index of the interval during which subject $i$ experienced the event or was censored, i.e., subject $i$ is at risk over interval $j \in \{1, \ldots J_i\}$, $\delta_{ij} = \nu_i \cdot 1\{y_i \in I_j\}$ equals $1$ if the $i^{th}$ individual experienced the event in interval $j$ and $0$ otherwise, and $f_{\text{Pois}}(z | \lambda) =  e^{-\lambda} \lambda^z / (z!)$ is the Poisson probability mass function (pmf) with mean $\lambda$. The last product in \eqref{eq:pweph_like} is equivalent to the likelihood of a Poisson GLM with $\theta_{ij} = \log \mu_{ij} = \log \lambda_j + \bm{x}_i'\bm{\beta} + \log R_j(y_i)$ for $i = 1, \ldots, n$, $j = 1, \ldots, J_i$. This corresponds to fitting a Poisson GLM with an interval-specific intercept and an offset given by the logarithm of the time at risk in interval $j$.

Using the Poisson representation of the likelihood in \eqref{eq:pweph_like}, we may fit the proportional hazards model in \textbf{hdbayes} by restructuring the data as follows. Let $\tilde{\bm{y}}_i = (\delta_{i,1}, \ldots, \delta_{i,J_i})'$ denote the $J_i$-dimensional vector containing the $\delta_{ij}'s$ for subject $i$ and let $\tilde{\bm{X}}_i = (\tilde{\bm{x}}_{i,1}, \ldots, \tilde{\bm{x}}_{i,J_i})'$, where $\tilde{\bm{x}}_{i,j} = (\bm{e}_j', \bm{x}_i')'$, $\bm{x}_i$ is a $p$-dimensional vector of covariates, and $\bm{e}_j = (e_{j1}, \ldots, e_{jJ})'$ is a $J$-dimensional vector with $e_{jk} = 1\{j=k\}$ for $k = 1, \ldots, J$, i.e., whose $j^{th}$ element is one and all other elements are zero. Also, let $\tilde{\bm{r}}_i = (r_{i,1}, \ldots, r_{i,J_i})'$ where $r_{i,j} = \log R_j(y_i)$ for $i = 1, \ldots, n$, $j = 1, \ldots, J_i$. Finally, let $\tilde{\bm{y}} = (\tilde{\bm{y}}_1', \ldots, \tilde{\bm{y}}_n')'$ denote a vector of size $N = \sum_{i=1}^n J_i$, $\tilde{\bm{X}} = (\tilde{\bm{X}}_1', \ldots, \tilde{\bm{X}}_n')'$ denote an $N \times (J + p)$ matrix, and $\tilde{\bm{r}} = ( \tilde{\bm{r}}_{1}', \ldots, \tilde{\bm{r}}_n')'$ denote an $N$-dimensional vector. The likelihood in \eqref{eq:pweph_like} may be expressed as the likelihood of a Poisson GLM with response vector $\tilde{\bm{y}}$, design matrix $\tilde{\bm{X}}$, and offset $\tilde{\bm{r}}$, giving the likelihood
\begin{align}
    L(\bm{\lambda}, \bm{\beta} | \tilde{\bm{y}}, \tilde{\bm{X}}, \tilde{\bm{r}})
    \propto
    \exp\left\{
        \tilde{\bm{y}}'\left[ \tilde{\bm{X}}\tilde{\bm{\beta}} + \tilde{\bm{r}} \right]
        - \bm{1}_N'\exp\left[ \tilde{\bm{X}}\tilde{\bm{\beta}} + \tilde{\bm{r}} \right]
    \right\}
    ,
    \label{eq:pwe_likelihood_matrix}
\end{align}
where $\tilde{\bm{\beta}} = ( [\log \bm{\lambda}]', \bm{\beta}')'$. The likelihood in \eqref{eq:pwe_likelihood_matrix} takes the form of a GLM likelihood \eqref{eq:glm_likelihood_matrix} with canonical link function $\theta(\bm{z}) = \bm{z}$, $b(\bm{\theta}) = e^{\bm{\theta}}$,  $\bm{c}(\bm{y}, \phi) = \bm{0}$, and an offset given by $\log \tilde{\bm{r}}$.

\section{Discussion}
\label{sec:discussion}
The package \textbf{hdbayes} provides functions with user-friendly, cohesive syntax to implement commonly used priors that incorporate historical data for generalized linear models. The full suite of GLMs and link functions available in the \textbf{stats} package is available in \textbf{hdbayes}, adding an unmet need in statistical software.

In the future, we aim to make several extensions to the \textbf{hdbayes} package. For example, more complex data types (e.g., longitudinal and time-to-event) could be included. Moreover, we aim to create more flexible types of borrowing, such as partial borrowing techniques and missing data. Finally, other types of historical data priors, such as propensity score integrated priors, could be included to provide a complete suite of historical data priors under a single package.

\newpage

\appendix

\setcounter{figure}{0}
\setcounter{table}{0}
\renewcommand{\thetable}{S\arabic{table}}
\renewcommand{\thefigure}{S\arabic{figure}}

\section*{ A. Link selection in binary regression with the power prior}
\label{sec:app_link_selection}

When employing generalized linear models, the choice of link function can be crucial, as it controls the degree of non-linearity between the conditional mean and the linear predictor. 
In this section we illustrate how to employ \textbf{hdbayes} to compute marginal likelihoods (and thus Bayes factors) under the power prior using the logit and probit links in binary regression.

Suppose $\boldsymbol{y}_0$ and $\boldsymbol{y}$ are the historical and current vectors of binary responses, respectively.
In keeping with the notation in the main text, we want to compare the logistic and probit models for the probability $\mu_i$ that $y_i = 1$:
\begin{align}
\mathcal{M}_1&: \mu_i = \frac{\exp\left(\bm{x}_i'\bm{\beta}\right)}{1 + \exp\left(\bm{x}_i'\bm{\beta}\right)},\\
\mathcal{M}_2&:  \mu_i = \Phi\left(\bm{x}_i'\bm{\beta}\right),
\end{align}
where $\Phi$ is the cumulative distribution function (CDF) of a standard normal random variable.
For simplicity, we use the same prior specification on the coefficients $\bm{\beta}$ for both models -- the default $\bm{\mu}_0 = \bm{0}_p$, $\bm{\sigma}_0 = 10 \cdot \bm{1}_p$ presented above.

We will use the power prior with fixed $a_0$ to elicit a prior for each model and employ bridge sampling to approximately compute the marginal likelihood
\begin{align}
    \label{eq:posterior_mal}
    m_i(a_0) &:=  \int_{\mathbb{R}^p} L_i(\bm{\beta} | D) \pi_{\text{PP}}(\bm{\beta} | D_{0}, a_{0}, \pi_0, \mathcal{M}_i) ~d\bm{\beta},\\
    \label{eq:prior_mal}
    \pi_{\text{PP}}(\bm{\beta} | D_{0}, a_{0}, \pi_0, \mathcal{M}_i) &:= \frac{L_i(\bm{\beta} | D_0)^{a_0}\pi_0(\bm{\beta})}{ \int_{\mathbb{R}^p} L_i(\bm{\beta} | D_0)^{a_0}\pi_0(\bm{\beta}) ~d\bm{\beta}},
\end{align}
where $D = (\bm{y}_0, \bm{x}_0)$ and $D = (\bm{y}, \bm{x})$ are the historical and current data sets, respectively, and $L_i$ and $\pi_{\text{PP}}(\cdot | \ldots, \mathcal{M}_i)$ are the likelihood and power prior under each model.
Notice that this computation necessitates two passes of the bridge sampling algorithm: one to estimate the power prior normalizing constant \eqref{eq:prior_mal} and another for the marginal likelihood of interest \eqref{eq:posterior_mal}.
The Bayes factor is then $\text{BF}_{12}(a_0) = m_1(a_0)/m_2(a_0)$.
It is interesting to understand how this quantity changes as one varies the discounting parameter, $a_0$.
Using the HIV data in Section~\hyperref[sec:dataanalysis_logistic]{4.1} we compute $m_i(a_0)$ for each model.

\begin{figure}[ht]
\centering
\includegraphics[scale=0.35]{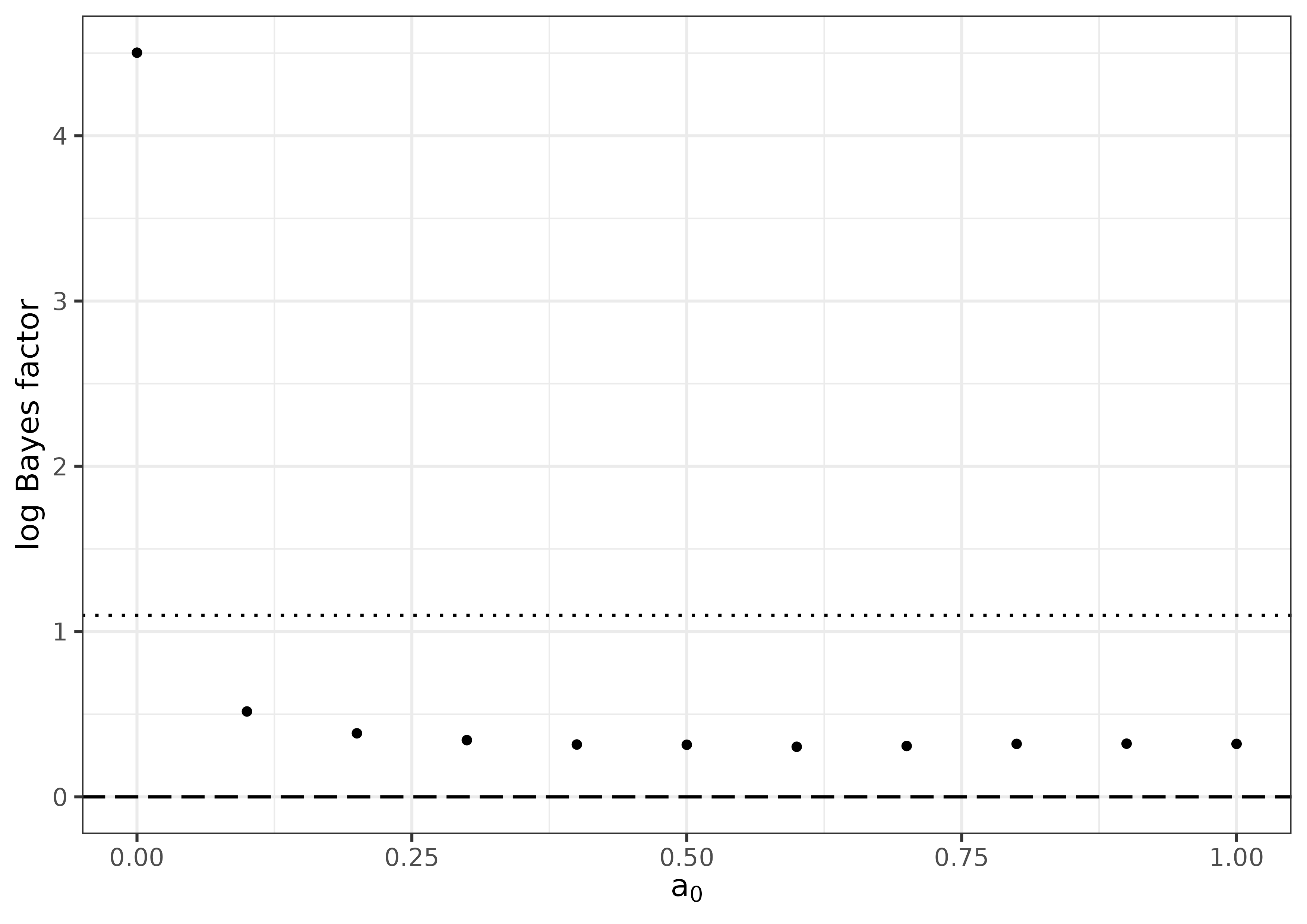}
\caption{\label{fig:link_selection}
\textbf{Log Bayes factor comparing logit (Model 1, $\mathcal{M}_1$) and probit ($\mathcal{M}_2$) for various values of the discounting parameter $a_0$}.
We show $\log\left(m_1(a_0)/m_2(a_0)\right)$.
Horizontal lines mark the $\log\text{BF}_{12} \geq 0$ (dashed) and $\log \text{BF}_{12} \geq \log(3) $ (dotted) thresholds.
}
\end{figure}

We show the results in Figure~\ref{fig:link_selection}, where the log of $\text{BF}_{12}(a_0)$ is plotted for a regular grid of values for the discounting parameter from $a_0 = 0$ to $a_0 = 1$ in increments of $0.1$ is shown.
Following~\cite{kass1995bayes}, we mark the $\text{BF}_{12}(a_0) \geq 1$ (dashed) and $\text{BF}_{12}(a_0) \geq 3$ (dotted) which correspond to weak and substantial evidence, respectively -- see Section 3.2 therein.
It is clear that the logit link is to be preferred over the probit model.
Interestingly, however, the relative support is not very strong, as evidenced by the fact that $\text{BF}_{12}(a_0) \geq 3$ only for $a_0 < 0.1$.
This suggests that one would need to introduce very little borrowing in order to observe strong support for the logit model.

\end{document}